# Efficiently Testing Sparse $GF(2)$ Polynomials

Ilias Diakonikolas, Homin K. Lee, Kevin Matulef,
Rocco A. Servedio, and Andrew Wan

{ilias,homin,rocco,atw12}@cs.columbia.edu, matulef@mit.edu

**Abstract.** We give the first algorithm that is both query-efficient and time-efficient for testing whether an unknown function $f : \{0,1\}^n \to \{0,1\}$ is an $s$-sparse $GF(2)$ polynomial versus $\epsilon$-far from every such polynomial. Our algorithm makes $\text{poly}(s, 1/\epsilon)$ black-box queries to $f$ and runs in time $n \cdot \text{poly}(s, 1/\epsilon)$. The only previous algorithm for this testing problem [DLM[+]07] used $\text{poly}(s, 1/\epsilon)$ queries, but had running time exponential in $s$ and super-polynomial in $1/\epsilon$.

Our approach significantly extends the "testing by implicit learning" methodology of [DLM[+]07]. The learning component of that earlier work was a brute-force exhaustive search over a concept class to find a hypothesis consistent with a sample of random examples. In this work, the learning component is a sophisticated exact learning algorithm for sparse $GF(2)$ polynomials due to Schapire and Sellie [SS96]. A crucial element of this work, which enables us to simulate the membership queries required by [SS96], is an analysis establishing new properties of how sparse $GF(2)$ polynomials simplify under certain restrictions of "low-influence" sets of variables.

## 1 Introduction

**Background and motivation.** Given black-box access to an unknown function $f : \{0,1\}^n \to \{0,1\}$, a natural question to ask is whether the function has a particular form. Is it representable by a small decision tree, or small circuit, or sparse polynomial? In the field of computational learning theory, the standard approach to this problem is to assume that $f$ belongs to a specific class $\mathcal{C}$ of functions of interest, and the goal is to identify or approximate $f$. In contrast, in property testing nothing is assumed about the unknown function $f$, and the goal of the testing algorithm is to output "yes" with high probability if $f \in \mathcal{C}$ and "no" with high probability if $f$ is $\epsilon$-far from every $g \in \mathcal{C}$. (Here the distance between two functions $f, g$ is measured with respect to the uniform distribution on $\{0,1\}^n$, so $f$ and $g$ are $\epsilon$-far if they disagree on more than an $\epsilon$ fraction of all inputs.) The complexity of a testing algorithm is measured both in terms of the number of black-box queries it makes to $f$ (*query complexity*) as well as the time it takes to process the results of those queries (*time complexity*).

There are many connections between learning theory and testing, and a growing body of work relating the two fields (see [Ron07] and its references). Testing algorithms have been given for a range of different function classes such as linear functions over $GF(2)$ (i.e. parities) [BLR93]; degree-$d$ $GF(2)$ polynomials [AKK[+]03]; Boolean literals, conjunctions, and $s$-term monotone DNF formulas [PRS02]; $k$-juntas (i.e. functions which depend on at most $k$ variables) [FKR[+]04]; halfspaces [MORS07]; and more.

Recently, Diakonikolas et al. [DLM[+]07] gave a general technique, called "testing by implicit learning," which they used to test a variety of different function classes that were not previously known to be testable. Intuitively, these classes correspond to functions with "concise representations," such as $s$-term DNFs, size-$s$ Boolean formulas, size-$s$ Boolean circuits, and $s$-sparse polynomials over constant-size finite fields. For each of these classes, the testing algorithm of [DLM$^+$07] makes only $\text{poly}(s, 1/\epsilon)$ queries (independent of $n$).

The main drawback of the [DLM$^+$07] testing algorithm is its time complexity. For each of the classes mentioned above, the algorithm's running time is $2^{\omega(s)}$ as a function of $s$, and $\omega(\text{poly}(1/\epsilon))$ as a function of $\epsilon$.[1] Thus, a natural question asked by [DLM$^+$07] is whether any of these classes can be tested with both time complexity and query complexity $\text{poly}(s, 1/\epsilon)$.

**Our result: efficiently testing sparse $GF(2)$ polynomials.** In this paper we focus on the class of $s$-sparse polynomials over $GF(2)$. Polynomials over $GF(2)$ (equivalently, parities of ANDs of input variables) are a simple and well-studied representation for Boolean functions. It is well known that every Boolean function has a unique representation as a multilinear polynomial over $GF(2)$, so the sparsity (number of monomials) of this polynomial is a very natural measure of the complexity of $f$. Sparse $GF(2)$ polynomials have been studied by many authors from a range of different perspectives such as learning [BS90,FS92,SS96,Bsh97a,BM02], approximation and interpolation [Kar89,GKS90,RB91], the complexity of (approximate) counting [EK89,KL93,LVW93], and property testing [DLM$^+$07].

The main result of this paper is a testing algorithm for $s$-sparse $GF(2)$ polynomials that is both time-efficient and query-efficient:

**Theorem 1.** *There is a $\text{poly}(s, 1/\epsilon)$-query algorithm with the following performance guarantee: given parameters $s, \epsilon$ and black-box access to any $f : \{0,1\}^n \to \{0,1\}$, it runs in time $\text{poly}(s, 1/\epsilon)$ and tests whether $f$ is an $s$-sparse $GF(2)$ polynomial versus $\epsilon$-far from every $s$-sparse polynomial.*

This answers the question of [DLM$^+$07] by exhibiting an interesting and natural class of functions with "concise representations" that can be tested efficiently, both in terms of query complexity and running time.

We obtain our main result by extending the "testing by implicit learning" approach of [DLM$^+$07]. In that work the "implicit learning" step used a naive brute-force search for a consistent hypothesis; in this paper we employ a sophisticated proper learning algorithm due to Schapire and Sellie [SS96]. It is much more difficult to "implicitly" run the [SS96] algorithm than the brute-force search of [DLM$^+$07]. One of the main technical contributions of this paper is a new structural theorem about how $s$-sparse $GF(2)$ polynomials are affected by certain carefully chosen restrictions; this is an essential ingredient that enables us to use the [SS96] algorithm. We elaborate on this below.

**Techniques.** We begin with a brief review of the main ideas of [DLM$^+$07]. The approach of [DLM$^+$07] builds on the observation of Goldreich et al. [GGR98] that any

---

[1] We note that the algorithm also has a linear running time dependence on $n$, the number of input variables; this is in some sense inevitable since the algorithm must set $n$ bit values just to pose a black-box query to $f$. Our algorithm has running time linear in $n$ for the same reason. For the rest of the paper we discuss the running time only as a function of $s$ and $\epsilon$.

*proper* learning algorithm for a function class $\mathcal{C}$ can be used as a testing algorithm for $\mathcal{C}$. (Recall that a proper learning algorithm for $\mathcal{C}$ is one which outputs a hypothesis $h$ that itself belongs to $\mathcal{C}$.) The idea behind this observation is that if the function $f$ being tested belongs to $\mathcal{C}$ then a proper learning algorithm will succeed in constructing a hypothesis that is close to $f$, while if $f$ is $\epsilon$-far from every $g \in \mathcal{C}$ then any hypothesis $h \in \mathcal{C}$ that the learning algorithm outputs must necessarily be far from $f$. Thus any class $\mathcal{C}$ can be tested to accuracy $\epsilon$ using essentially the same number of queries that are required to properly learn the class to accuracy $\Theta(\epsilon)$.

The basic approach of [GGR98] did not yield query-efficient testing algorithms (with query complexity independent of $n$) since virtually every interesting class of functions over $\{0,1\}^n$ requires $\Omega(\log n)$ examples for proper learning. However, [DLM$^+$07] showed that for many classes of functions defined by a size parameter $s$, it is possible to "implicitly" run a (very naive) proper learning algorithm over a number of variables that is independent of $n$, and thus obtain an overall query complexity independent of $n$. More precisely, they first observed that for many classes $\mathcal{C}$ every $f \in \mathcal{C}$ is "very close" to a function $f' \in \mathcal{C}$ for which the number $r$ of relevant variables is polynomial in $s$ and independent of $n$; roughly speaking, the relevant variables for $f'$ are the variables that have high influence in $f$. (For example, if $f$ is an $s$-sparse $GF(2)$ polynomial, an easy argument shows that there is a function $f'$ – obtained by discarding from $f$ all monomials of degree more than $\log(s/\tau)$ – that is $\tau$-close to $f$ and depends on at most $r = s \log(s/\tau)$ variables.) They then showed how, using ideas of Fischer et al. [FKR$^+$04] for testing juntas, it is possible to construct a sample of uniform random examples over $\{0,1\}^r$ which with high probability are all labeled according to $f'$. At this point, the proper learning algorithm employed by [DLM$^+$07] was a naive brute-force search. The algorithm tried all possible functions in $\mathcal{C}$ over $r$ (as opposed to $n$) variables, to see if any were consistent with the labeled sample. [DLM$^+$07] thus obtained a testing algorithm with overall query complexity poly$(s/\epsilon)$ but whose running time was dominated by the brute-force search. For the class of $s$-sparse $GF(2)$ polynomials, their algorithm used $\tilde{O}(s^4/\epsilon^2)$ queries but had running time at least $2^{\omega(s)} \cdot (1/\epsilon)^{\log \log(1/\epsilon)}$.

**Current approach.** The high-level idea of the current work is to employ a much more sophisticated – and efficient – proper learning algorithm than brute-force search. In particular we would like to use a proper learning algorithm which, when applied to learn a function over only $r$ variables, runs in time polynomial in $r$ and in the size parameter $s$. For the class of $s$-sparse $GF(2)$ polynomials, precisely such an algorithm was given by Schapire and Sellie [SS96]. Their algorithm, which we describe in Section 2.1, is computationally efficient and generates a hypothesis $h$ which is an $s$-sparse $GF(2)$ polynomial. But this power comes at a price: the algorithm requires access to a *membership query* oracle, i.e. a black-box oracle for the function being learned. Thus, in order to run the Schapire/Sellie algorithm in the "testing by implicit learning" framework, it is necessary to simulate membership queries to an approximating function $f' \in \mathcal{C}$ which is close to $f$ but depends on only $r$ variables. This is significantly more challenging than generating uniform random examples labeled according to $f'$, which is all that is required in the original [DLM$^+$07] approach.

To see why membership queries to $f'$ are more difficult to simulate than uniform random examples, recall that $f$ and the $f'$ described above (obtained from $f$ by discard-ing high-degree monomials) are $\tau$-close. Intuitively this is extremely close, disagreeing only on a $1/m$ fraction of inputs for an $m$ that is much larger than the number of random examples required for learning $f'$ via brute-force search (this number is "small" – independent of $n$ – because $f'$ depends on only $r$ variables). Thus in the [DLM[+]07] approach it suffices to use $f$, the function to which we actually have black-box access, rather than $f'$ to label the random examples used for learning $f'$; since $f$ and $f'$ are so close, and the examples are uniformly random, with high probability all the labels will also be correct for $f'$. However, in the membership query scenario of the current paper, things are no longer that simple. For any given $f'$ which is close to $f$, one can no longer assume that the learning algorithm's queries to $f'$ are uniformly distributed and hence unlikely to hit the error region – indeed, it is possible that the learning algorithm's membership queries to $f'$ are clustered on the few inputs where $f$ and $f'$ disagree.

In order to successfully simulate membership queries, we must somehow consistently answer queries according to a particular $f'$, even though we only have oracle access to $f$. Moreover this must be done implicitly in a query-efficient way, since explicitly identifying even a single variable relevant to $f'$ requires at least $\Omega(\log n)$ queries. This is the main technical challenge in the paper.

We meet this challenge by showing that for any $s$-sparse polynomial $f$, an approximating $f'$ can be obtained as a restriction of $f$ by setting certain carefully chosen subsets of variables to zero. Roughly speaking, this restriction is obtained by randomly partitioning all of the input variables into $r$ subsets and zeroing out all subsets whose variables have small "collective influence" (more precisely, small variation in the sense of [FKR[+]04]). It is important that the restriction sets these variables to zero, rather than a random assignment; intuitively this is because setting a variable to zero "kills" all monomials that contain the variable, whereas setting it to 1 does not. Our main technical theorem (Theorem 3, given in Section 3) shows that this $f'$ is indeed close to $f$ and has at most one of its relevant variables in each of the surviving subsets. We moreover show that these relevant variables for $f'$ all have high influence in $f$ (the converse is not true; examples can be given which show that not every variable that has "high influence" in $f$ will in general become a relevant variable for $f'$). This property is important in enabling our simulation of membership queries. In addition to the crucial role that Theorem 3 plays in the completeness proof for our test, we feel that the new insights the theorem gives into how sparse polynomials "simplify" under (appropriately defined) random restrictions may be of independent interest.

**Organization.** In Section 4, we present our testing algorithm, **Test-Sparse-Poly**, along with a high-level description and sketch of correctness. In Section 2.1 we describe in detail the "learning component" of the algorithm. In Section 3 we state Theorem 3, which provides intuition behind the algorithm and serves as the main technical tool in the completeness proof. Due to space limitations, the proof of Theorem 3 is presented in Appendix A, while the completeness and soundness proofs are given in Appendices B and C, respectively (see full version available online).

## 2   Preliminaries and Background

*GF*(**2**) **Polynomials:** A $GF(2)$ polynomial is a parity of monotone conjunctions (monomials). It is *s-sparse* if it contains at most $s$ monomials (including the constant-1 mono-mial if it is present). The *length* of a monomial is the number of distinct variables that occur in it; over $GF(2)$, this is simply its degree.

*Notation:* For $i \in \mathbb{N}^*$, denote $[i] \stackrel{\text{def}}{=} \{1, 2, \ldots, i\}$. It will be convenient to view the output range of a Boolean function $f$ as $\{-1, 1\}$ rather than $\{0, 1\}$, i.e. $f : \{0, 1\}^n \to \{-1, 1\}$. We view the hypercube as a measure space endowed with the uniform product probability measure. For $I \subseteq [n]$ we denote by $\{0, 1\}^I$ the set of all partial assignments to the coordinates in $I$. For $w \in \{0, 1\}^{[n] \setminus I}$ and $z \in \{0, 1\}^I$, we write $w \sqcup z$ to denote the assignment in $\{0, 1\}^n$ whose $i$-th coordinate is $w_i$ if $i \in [n] \setminus I$ and is $z_i$ if $i \in I$. Whenever an element $z$ in $\{0, 1\}^I$ is chosen randomly (we denote $z \in_R \{0, 1\}^I$), it is chosen with respect to the uniform measure on $\{0, 1\}^I$.

**Influence, Variation and the Independence Test:** Recall the classical notion of *influence* [KKL88]: The *influence* of the $i$-th coordinate on $f : \{0,1\}^n \to \{-1, 1\}$ is $\text{Inf}_i(f) \stackrel{\text{def}}{=} \Pr_{x \in_R \{0,1\}^n}[f(x) \neq f(x^{\oplus i})]$, where $x^{\oplus i}$ denotes $x$ with the $i$-th bit flipped. The following generalization of influence, the *variation* of a subset of the coordinates of a Boolean function, plays an important role for us:

**Definition 1 (variation, [FKR+04]).** *Let $f : \{0,1\}^n \to \{-1, 1\}$, and let $I \subseteq [n]$. We define the* variation *of $f$ on $I$ as $\text{Vr}_f(I) \stackrel{\text{def}}{=} \mathbb{E}_{w \in_R \{0,1\}^{[n] \setminus I}} \left[ \mathbb{V}_{z \in_R \{0,1\}^I} [f(w \sqcup z)] \right].$*

When $I = \{i\}$ we will sometimes write $\text{Vr}_f(i)$ instead of $\text{Vr}_f(\{i\})$. It is easy to check that $\text{Vr}_f(i) = \text{Inf}_i(f)$, so variation is indeed a generalization of influence. Intuitively, the variation is a measure of the ability of a set of variables to sway a function's output. The following two simple properties of the variation will be useful for the analysis of our testing algorithm:

**Lemma 1 (monotonicity and sub-additivity, [FKR+04]).** *Let $f : \{0,1\}^n \to \{-1, 1\}$ and $A, B \subseteq [n]$. Then $\text{Vr}_f(A) \leq \text{Vr}_f(A \cup B) \leq \text{Vr}_f(A) + \text{Vr}_f(B)$.*

**Lemma 2 (probability of detection, [FKR+04]).** *Let $f : \{0,1\}^n \to \{-1, 1\}$ and $I \subseteq [n]$. If $w \in_R \{0,1\}^{[n] \setminus I}$ and $z_1, z_2 \in_R \{0,1\}^I$ are chosen independently, then $\Pr[f(w \sqcup z_1) \neq f(w \sqcup z_2)] = \frac{1}{2} \text{Vr}_f(I)$.*

We now recall the *independence test* from [FKR+04], a simple two query test used to determine whether a function $f$ is independent of a given set $I \subseteq [n]$ of coordinates.

**Independence test:** Given $f : \{0, 1\}^n \to \{-1, 1\}$ and $I \subseteq [n]$, choose $w \in_R \{0,1\}^{[n] \setminus I}$ and $z_1, z_2 \in_R \{0,1\}^I$ independently. Accept if $f(w \sqcup z_1) = f(w \sqcup z_2)$ and reject if $f(w \sqcup z_1) \neq f(w \sqcup z_2)$.

Lemma 2 implies that the independence test rejects with probability exactly $\frac{1}{2}\text{Vr}_f(I)$.

**Random Partitions:** Throughout the paper we will use the following notion of a random partition of the set $[n]$ of input coordinates:

**Definition 2.** *A random partition of $[n]$ into $r$ subsets $\{I_j\}_{j=1}^r$ is constructed by independently assigning each $i \in [n]$ to a randomly chosen $I_j$ for some $j \in [r]$.*

We now define the notion of low- and high-variation subsets with respect to a partition of the set $[n]$ and a parameter $\alpha > 0$.

**Definition 3.** *For $f : \{0,1\}^n \to \{-1,1\}$, a partition of $[n]$ into $\{I_j\}_{j=1}^r$ and a parameter $\alpha > 0$, define $L(\alpha) \stackrel{def}{=} \{j \in [r] \mid \mathrm{Vr}_f(I_j) < \alpha\}$ (low-variation subsets) and $H(\alpha) \stackrel{def}{=} [r] \setminus L(\alpha)$ (high-variation subsets). For $j \in [r]$ and $i \in I_j$, if $\mathrm{Vr}_f(i) \geq \alpha$ we say that the variable $x_i$ is a* high-variation element *of $I_j$.*

Finally, the notion of a *well-structured* subset will be important for us:

**Definition 4.** *For $f : \{0,1\}^n \to \{-1,1\}$ and parameters $\alpha > \Delta > 0$, we say that a subset $I \subseteq [n]$ of coordinates is $(\alpha, \Delta)$-well structured if there is an $i \in I$ such that $\mathrm{Vr}_f(i) \geq \alpha$ and $\mathrm{Vr}_f(I \setminus \{i\}) \leq \Delta$.*

Note that since $\alpha > \Delta$, by monotonicity, the $i \in I$ in the above definition is unique. Hence, a well-structured subset contains a single high-influence coordinate, while the remaining coordinates have small total variation.

### 2.1 Background on Schapire and Sellie's algorithm.

In [SS96] Schapire and Sellie gave an algorithm, which we refer to as **LearnPoly**, for exactly learning $s$-sparse $GF(2)$ polynomials using membership queries (i.e. black-box queries) and equivalence queries. Their algorithm is *proper*; this means that every equivalence query the algorithm makes (including the final hypothesis of the algorithm) is an $s$-sparse polynomial. (We shall see that it is indeed crucial for our purposes that the algorithm is proper.) Recall that in an equivalence query the learning algorithm proposes a hypothesis $h$ to the oracle: if $h$ is logically equivalent to the target function being learned then the response is "correct" and learning ends successfully, otherwise the response is "no" and the learner is given a counterexample $x$ such that $h(x) \neq f(x)$.

Schapire and Sellie proved the following about their algorithm:

**Theorem 2.** *[[SS96], Theorem 10] Algorithm **LearnPoly** is a proper exact learning algorithm for the class of $s$-sparse $GF(2)$ polynomials over $\{0,1\}^n$. The algorithm runs in $\mathrm{poly}(n,s)$ time and makes at most $\mathrm{poly}(n,s)$ membership queries and at most $ns + 2$ equivalence queries.*

We can easily also characterize the behavior of **LearnPoly** if it is run on a function $f$ that is not an $s$-sparse polynomial. In this case, since the algorithm is proper all of its equivalence queries have $s$-sparse polynomials as their hypotheses, and consequently no equivalence query will ever be answered "correct." So if the $(ns+2)$-th equivalence query is not answered "correct," the algorithm may infer that the target function is not an $s$-sparse polynomial, and it returns "not $s$-sparse."

A well-known result due to Angluin [Ang88] says that in a Probably Approximately Correct or PAC setting (where there is a distribution $\mathcal{D}$ over examples and the goal is to construct an $\epsilon$-accurate hypothesis with respect to that distribution), equivalence queries can be straightforwardly simulated using random examples. This is done simply by drawing a sufficiently large sample of random examples for each equivalence query and evaluting both the hypothesis $h$ and the target function $f$ on each point in the sample. This either yields a counterexample (which simulates an equivalence query), or if no counterexample is obtained then simple arguments show that for a large enough ($O(\log(1/\delta)/\epsilon)$-size) sample, with probability $1 - \delta$ the functions $f$ and $h$ must be $\epsilon$-close under the distribution $\mathcal{D}$, which is the success criterion for PAC learning. This directly gives the following corollary of Theorem 2:

**Corollary 1.** *There is a uniform distribution membership query proper learning algorithm, which we call* **LearnPoly**$'(s, n, \epsilon, \delta)$, *which makes* $Q(s, n, \epsilon, \delta) \stackrel{\text{def}}{=} \text{poly}(s, n, 1/\epsilon, \log(1/\delta))$ *membership queries and runs in* $\text{poly}(Q)$ *time to learn $s$-sparse polynomials over $\{0, 1\}^n$ to accuracy $\epsilon$ and confidence $1 - \delta$ under the uniform distribution.*

## 3  On restrictions which simplify sparse polynomials

This section presents Theorem 3, which gives the intuition behind our testing algorithm, and lies at the heart of the completeness proof. We give the full proof of Theorem 3 in Appendix A (see the full version).

Roughly speaking, the theorem says the following: consider any $s$-sparse $GF(2)$ polynomial $p$. Suppose that its coordinates are randomly partitioned into $r = \text{poly}(s)$ many subsets $\{I_j\}_{j=1}^r$. The first two statements say that w.h.p. a randomly chosen "threshold value" $\alpha \approx 1/\text{poly}(s)$ will have the property that no single coordinate $i$, $i \in [n]$, or subset $I_j$, $j \in [r]$, has $\text{Vr}_p(i)$ or $\text{Vr}_p(I_j)$ "too close" to $\alpha$. Moreover, the high-variation subsets (w.r.t. $\alpha$) are precisely those that contain a single high variation element $i$ (i.e. $\text{Vr}_p(i) \geq \alpha$), and in fact each such subset $I_j$ is well-structured (part 3). Also, the number of such high-variation subsets is small (part 4). Finally, let $p'$ be the restriction of $p$ obtained by setting all variables in the low-variation subsets to 0. Then, $p'$ has a nice structure: it has at most one relevant variable per high-variation subset (part 5), and it is close to $p$ (part 6).

**Theorem 3.** *Let $p : \{0, 1\}^n \to \{-1, 1\}$ be an $s$-sparse polynomial. Fix $\tau \in (0, 1)$ and $\Delta$ such that $\Delta \leq \Delta_0 \stackrel{\text{def}}{=} \tau/(1600 s^3 \log(8s^3/\tau))$ and $\Delta = \text{poly}(\tau/s)$. Let $r \stackrel{\text{def}}{=} 4Cs/\Delta$, for a suitably large constant $C$. Let $\{I_j\}_{j=1}^r$ be a random partition of $[n]$. Choose $\alpha$ uniformly at random from the set $\mathcal{A}(\tau, \Delta) \stackrel{\text{def}}{=} \{\frac{\tau}{4s^2} + (8\ell - 4)\Delta : \ell \in [K]\}$ where $K$ is the largest integer such that $8K\Delta \leq \frac{\tau}{4s^2}$. Then with probability at least $9/10$ (over the choice of $\alpha$ and $\{I_j\}_{j=1}^r$), all of the following statements hold:*

1. *Every variable $x_i$, $i \in [n]$, has $\text{Vr}_p(i) \notin [\alpha - 4\Delta, \alpha + 4\Delta]$.*
2. *Every subset $I_j$, $j \in [r]$, has $\text{Vr}_p(I_j) \notin [\alpha - 3\Delta, \alpha + 4\Delta]$.*
3. *For every $j \in H(\alpha)$, $I_j$ is $(\alpha, \Delta)$-well structured.*
4. *$|H(\alpha)| \leq s \log(8s^3/\tau)$.*

*Let $p' \stackrel{\text{def}}{=} p|_{0 \leftarrow \cup_{j \in L(\alpha)} I_j}$ (the restriction obtained by fixing all variables in low-variation subsets to 0).*

5. *For every $j \in H(\alpha)$, $p'$ has at most one relevant variable in $I_j$ (hence $p'$ is a $|H(\alpha)|$-junta).*

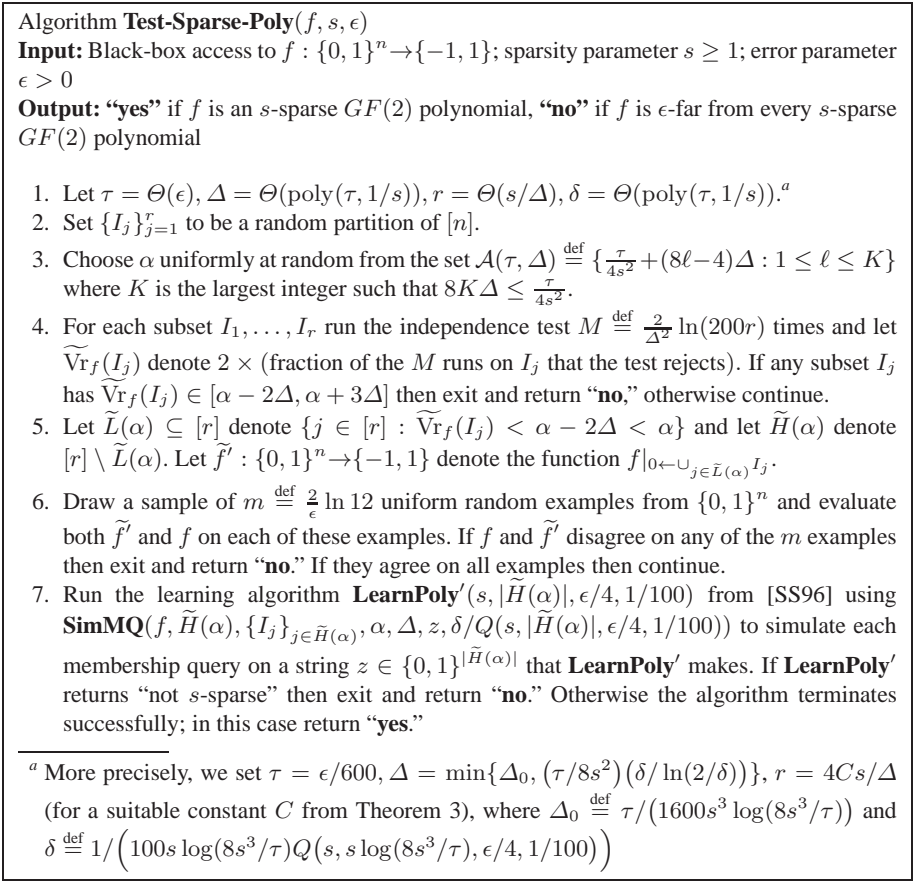

**Fig. 1.** The algorithm **Test-Sparse-Poly**.

6. *The function $p'$ is $\tau$-close to $p$.*

Theorem 3 naturally suggests a testing algorithm, whereby we attempt to partition the coordinates of a function $f$ into "high-variation" subsets and "low-variation" subsets, then zero-out the variables in low-variation subsets and implicitly learn the remaining function $f'$ on only $\text{poly}(s, 1/\epsilon)$ many variables. This is exactly the approach we take in the next section.

## 4 The testing algorithm Test-Sparse-Poly

In this section we present our main testing algorithm and give high-level sketches of the arguments establishing its completeness and soundness. The algorithm, which is called **Test-Sparse-Poly**, takes as input the values $s, \epsilon > 0$ and black-box access to $f : \{0,1\}^n \to \{-1,1\}$. It is presented in full in Figure 1.

---

Algorithm **Set-High-Influence-Variable**$(f, I, \alpha, \Delta, b, \delta)$
**Input:** Black-box access to $f : \{0,1\}^n \to \{-1,1\}$; $(\alpha, \Delta)$-well-structured set $I \subseteq [n]$; bit $b \in \{0, 1\}$; failure parameter $\delta$.
**Output:** assignment $w \in \{0,1\}^I$ to the variables in $I$ such that $w_i = b$ with probability $1 - \delta$

1. Draw $x$ uniformly from $\{0,1\}^I$. Define $I^0 \stackrel{\text{def}}{=} \{j \in I : x_j = 0\}$ and $I^1 \stackrel{\text{def}}{=} \{j \in I : x_j = 1\}$.
2. Apply $c = \frac{2}{\alpha} \ln(\frac{2}{\delta})$ iterations of the *independence test* to $(f, I^0)$. If any of the $c$ iterations reject, mark $I^0$. Do the same for $(f, I^1)$.
3. If both or neither of $I^0$ and $I^1$ are marked, stop and output "fail".
4. If $I^b$ is marked then return the assignment $w = x$. Otherwise return the assignment $w = \overline{x}$ (the bitwise negation of $x$).

---

**Fig. 2.** The subroutine **Set-High-Influence-Variable**.

---

Algorithm **SimMQ**$(f, H, \{I_j\}_{j \in H}, \alpha, \Delta, z, \delta)$
**Input:** Black-box access to $f : \{0,1\}^n \to \{-1,1\}$; subset $H \subseteq [r]$; disjoint subsets $\{I_j\}_{j \in H}$ of $[n]$; parameters $\alpha > \Delta$; string $z \in \{0,1\}^{|H|}$; failure probability $\delta$
**Output:** bit $b$ which, with probability $1 - \delta$ is the value of $f'$ on a random assignment $x$ in which each high-variation variable $i \in I_j$ ($j \in H$) is set according to $z$

1. For each $j \in H$, call **Set-High-Influence-Variable**$(f, I_j, \alpha, \Delta, z_j, \delta/|H|)$ and get back an assignment (call it $w^j$) to the variables in $I_j$.
2. Construct $x \in \{0,1\}^n$ as follows: for each $j \in H$, set the variables in $I_j$ according to $w^j$. This defines $x_i$ for all $i \in \cup_{j \in H} I_j$. Set $x_i = 0$ for all other $i \in [n]$.
3. Return $b = f(x)$.

---

**Fig. 3.** The subroutine **SimMQ**.

The first thing **Test-Sparse-Poly** does (Step 2) is randomly partition the coordinates into $r = \tilde{O}(s^4/\tau)$ subsets. In Steps 3 and 4 the algorithm attempts to distinguish subsets that contain a high-influence variable from subsets that do not; this is done by using the independence test to estimate the variation of each subset (see Lemma 2).

Once the high-variation and low-variation subsets have been identified, intuitively we would like to focus our attention on the high-influence variables. Thus, Step 5 of the algorithm defines a function $\widetilde{f'}$ which "zeroes out" all of the variables in all low-variation subsets. Step 6 of **Test-Sparse-Poly** checks that $f$ is close to $\widetilde{f'}$

The final step of **Test-Sparse-Poly** is to run the algorithm **LearnPoly'** of [SS96] to learn a sparse polynomial, which we call $\widetilde{f''}$, which is isomorphic to $\widetilde{f'}$ but is defined only over the high-influence variables of $f$ (recall that if $f$ is indeed $s$-sparse, there is at most one from each high-variation subset). The overall **Test-Sparse-Poly** algorithm accepts $f$ if and only if **LearnPoly'** successfully returns a final hypothesis (i.e. does not halt and output "fail"). The membership queries that the [SS96] algorithm requires are simulated using the **SimMQ** procedure, which in turn uses a subroutine called **Set-High-Influence-Variables**.

The procedure **Set-High-Influence-Variable** (**SHIV**) is presented in Figure 2. The idea of this procedure is that when it is run on a well-structured subset of variables $I$, it returns an assignment in which the high-variation variable is set to the desired bit value. Intuitively, the executions of the independence test in the procedure are used to determine whether the high-variation variable $i \in I$ is set to 0 or 1 under the assignment $x$. Depending on whether this setting agrees with the desired value, the algorithm either returns $x$ or the bitwise negation of $x$ (this is slightly different from **Construct-Sample**, the analogous subroutine in [DLM+07], which is content with a random $x$ and thus never needs to negate coordinates).

Figure 3 gives the **SimMQ** procedure. When run on a function $f$ and a collection $\{I_j\}_{j \in H}$ of disjoint well-structured subsets of variables, **SimMQ** takes as input a string $z$ of length $|H|$ which specifies a desired setting for each high-variation variable in each $I_j$ ($j \in H$). **SimMQ** constructs a random assignment $x \in \{0,1\}^n$ such that the high-variation variable in each $I_j$ ($j \in H$) is set in the desired way in $x$, and it returns the value $f'(x)$.

### 4.1 Time and Query Complexity of Test-Sparse-Poly

As stated in Figure 1, the **Test-Sparse-Poly** algorithm runs **LearnPoly**$'(s, |\widetilde{H}(\alpha)|, \epsilon/4, 1/100)$ using **SimMQ**$(f, \widetilde{H}(\alpha), \{I_j\}_{j \in \widetilde{H}(\alpha)}, \alpha, \Delta, z, 1/(100Q(s, |\widetilde{H}(\alpha)|, z, 1/100)))$ to simulate each membership query on an input string $z \in \{0,1\}^{|\widetilde{H}(\alpha)|}$. Thus the algorithm is being run over a domain of $|\widetilde{H}(\alpha)|$ variables. Since we certainly have $|\widetilde{H}(\alpha)| \leq r \leq \text{poly}(s, \frac{1}{\epsilon})$, Corollary 1 gives that **LearnPoly**$'$ makes at most $\text{poly}(s, \frac{1}{\epsilon})$ many calls to **SimMQ**. From this point, by inspection of **SimMQ**, **SHIV** and **Test-Sparse-Poly**, it is straightforward to verify that **Test-Sparse-Poly** indeed makes $\text{poly}(s, \frac{1}{\epsilon})$ many queries to $f$ and runs in time $\text{poly}(s, \frac{1}{\epsilon})$ as claimed in Theorem 1. Thus, to prove Theorem 1 it remains only to establish correctness of the test.

### 4.2 Sketch of completeness

The main tool behind our completeness argument is Theorem 3. Suppose $f$ is indeed an $s$-sparse polynomial. Then Theorem 3 guarantees that a randomly chosen $\alpha$ will w.h.p. yield a "gap" such that subsets with a high-influence variable have variation above the gap, and subsets with no high-influence variable have variation below the gap. This means that the estimates of each subset's variation (obtained by the algorithm in step 4) are accurate enough to effectively separate the high-variation subsets from the low-variation ones in step 5. Thus, the function $\widetilde{f'}$ defined by the algorithm will w.h.p be equal to the function $p'$ from Theorem 3.

Assuming that $f$ is an $s$-sparse polynomial (and that $\widetilde{f'}$ is equal to $p'$), Theorem 3 additionally implies that the function $\widetilde{f'}$ will be close to the original function (so Step 6 will pass), that $\widetilde{f'}$ only depends on $\text{poly}(s, 1/\epsilon)$ many variables, and that all of the subsets $I_j$ that "survive" into $\widetilde{f'}$ are well-structured. As we show in Appendix B, this condition is sufficient to ensure that **SimMQ** can successfully simulate membership queries to $\widetilde{f''}$. Thus, for $f$ an $s$-sparse polynomial, the **LearnPoly**$'$ algorithm can run successfully, and the test will accept.

### 4.3 Sketch of soundness

Here, we briefly argue that if **Test-Sparse-Poly** accepts $f$ with high probability, then $f$ must be close to some $s$-sparse polynomial (we give the full proof in Appendix C). Note that if $f$ passes Step 4, then **Test-Sparse-Poly** must have obtained a partition of variables into "high-variation" subsets and "low-variation" subsets. If $f$ passes Step 6, then it must moreover be the case that $f$ is close to the function $\widetilde{f}'$ obtained by zeroing out the low-variation subsets.

In the last step, **Test-Sparse-Poly** attempts to run the **LearnPoly**$'$ algorithm using $\widetilde{f}'$ and the high-variation subsets; in the course of doing this, it makes calls to **SimMQ**. Since $f$ could be an arbitrary function, we do not know whether each high-variation subset has at most one variable relevant to $\widetilde{f}'$ (as would be the case, by Theorem 3, if $f$ were an $s$-sparse polynomial). However, we are able to show (Lemma 11) that, if with high probability all calls to the **SimMQ** routine are answered without its ever returning "fail," then $\widetilde{f}'$ must be close to a junta $g$ whose relevant variables are the individual "highest-influence" variables in each of the high-variation subsets. Now, given that **LearnPoly**$'$ halts successfully, it must be the case that it constructs a final hypothesis $h$ that is itself an $s$-sparse polynomial and that agrees with many calls to **SimMQ** on random examples. Lemma 12 states that, in this event, $h$ must be close to $g$, hence close to $\widetilde{f}'$, and hence close to $f$.

## 5 Conclusion and future directions

An obvious question raised by our work is whether similar methods can be used to efficiently test $s$-sparse polynomials over a general finite field $\mathbb{F}$, with query and time complexity polynomial in $s$, $1/\epsilon$, and $|\mathbb{F}|$. The basic algorithm of [DLM$^+$07] uses $\tilde{O}((s|\mathbb{F}|)^4/\epsilon^2)$ queries to test $s$-sparse polynomials over $\mathbb{F}$, but has running time $2^{\omega(s|\mathbb{F}|)} \cdot (1/\epsilon)^{\log \log(1/\epsilon)}$ (arising, as discussed in Section 1, from brute-force search for a consistent hypothesis.). One might hope to improve that algorithm by using techniques from the current paper. However, doing so requires an algorithm for properly learning $s$-sparse polynomials over general finite fields. To the best of our knowledge, the most efficient algorithm for doing this (given only black-box access to $f : \mathbb{F}^n \to \mathbb{F}$) is the algorithm of Bshouty [Bsh97b] which requires $m = s^{O(|\mathbb{F}| \log |\mathbb{F}|)} \log n$ queries and runs in $\mathrm{poly}(m, n)$ time. (Other learning algorithms are known which do not have this exponential dependence on $|\mathbb{F}|$, but they either require evaluating the polynomial at complex roots of unity [Man95] or on inputs belonging to an extension field of $\mathbb{F}$ [GKS90,Kar89].) It would be interesting to know whether there is a testing algorithm that simultaneously achieves a polynomial runtime (and hence query complexity) dependence on both the size parameter $s$ and the cardinality of the field $|\mathbb{F}|$.

Another goal for future work is to apply our methods to other classes beyond just polynomials. Is it possible to combine the "testing by implicit learning" approach of [DLM$^+$07] with other membership-query-based learning algorithms, to achieve time and query efficient testers for other natural classes?

# A  Proof of Theorem 3

In Section A.1 we prove some useful preliminary lemmas about the variation of individual variables in sparse polynomials. In Section A.2 we extend this analysis to get high-probability statements about variation of subsets $\{I_j\}_{j=1}^r$ in a random partition. We put the pieces together to finish the proof of Theorem 3 in Section A.3.

Throughout this section the parameters $\tau$, $\Delta$, $r$ and $\alpha$ are all as defined in Theorem 3.

## A.1  The influence of variables in $s$-sparse polynomials

We start with a simple lemma stating that only a small number of variables can have large variation:

**Lemma 3.** *Let $p : \{0,1\}^n \to \{-1,1\}$ be an $s$-sparse polynomial. For any $\delta > 0$, there are at most $s\log(2s/\delta)$ many variables $x_i$ that have $\mathrm{Vr}_p(i) \geq \delta$.*

*Proof.* Any variable $x_i$ with $\mathrm{Vr}_p(i) \geq \delta$ must occur in some term of length at most $\log(2s/\delta)$. (Otherwise each occurrence of $x_i$ would contribute less than $\delta/s$ to the variation of the $i$-th coordinate, and since there are at most $s$ terms this would imply $\mathrm{Vr}_p(i) < s \cdot (\delta/s) = \delta$.) Since at most $s\log(2s/\delta)$ distinct variables can occur in terms of length at most $\log(2s/\delta)$, the lemma follows. ∎

**Lemma 4.** *With probability at least $96/100$ over the choice of $\alpha$, no variable $x_i$ has $\mathrm{Vr}_p(i) \in [\alpha - 4\Delta, \alpha + 4\Delta]$.*

*Proof.* The uniform random variable $\alpha$ has support $\mathcal{A}(\tau, \Delta)$ of size no less than $50s\log(8s^3/\tau)$. Each possible value of $\alpha$ defines the interval of variations $[\alpha - 4\Delta, \alpha + 4\Delta]$. Note that $\alpha - 4\Delta \geq \tau/(4s^2)$. In other words, the only variables which could lie in $[\alpha - 4\Delta, \alpha + 4\Delta]$ are those with variation at least $\tau/(4s^2)$. By Lemma 3 there are at most $k \stackrel{\mathrm{def}}{=} s\log(8s^3/\tau)$ such candidate variables. Since we have at least $50k$ intervals (two consecutive such intervals overlap at a single point) and at most $k$ candidate variables, by the pigeonhole principle, at least $48k$ intervals will be empty. ∎

Lemma 3 is based on the observation that, in a sparse polynomial, a variable with "high" influence (variation) must occur in some "short" term. The following lemma is in some sense a quantitative converse: it states that a variable with "small" influence can only appear in "long" terms.

**Lemma 5.** *Let $p : \{0,1\}^n \to \{-1,1\}$ be an $s$-sparse polynomial. Suppose that $i$ is such that $\mathrm{Vr}_p(i) < \tau/(s^2 + s)$. Then the variable $x_i$ appears only in terms of length greater than $\log(s/\tau)$.*

*Proof.* By contradiction. Assuming that $x_i$ appears in some term of length at most $\log(s/\tau)$, we will show that $\mathrm{Vr}_p(i) \geq \tau/(s^2 + s)$. Let $T$ be a shortest term that $x_i$ appears in. The function $p$ can be uniquely decomposed as follows: $p(x_1, x_2, \ldots, x_n) = x_i \cdot (T' + p_1) + p_2$, where $T = x_i \cdot T'$, the term $T'$ has length less than $\log(s/\tau)$ and does not depend on $x_i$, and $p_1$, $p_2$ are $s$-sparse polynomials that do not depend on $x_i$. Observe that since $T$ is a shortest term that contains $x_i$, the polynomial $p_1$ does not contain the constant term 1.

Since $T'$ contains fewer than $\log(s/\tau)$ many variables, it evaluates to 1 on at least a $\tau/s$ fraction of all inputs. The partial assignment that sets all the variables in $T'$ to 1 induces an $s$-sparse polynomial $p'_1$ (the restriction of $p_1$ according to the partial assignment). Now observe that $p'_1$ still does not contain the constant term 1 (for since each term in $p_1$ is of length at least the length of $T'$, no term in $p_1$ is a subset of the variables in $T'$). We now recall the following (nontrivial) result of Karpinski and Luby [KL93]:

*Claim ([KL93], Corollary 1).* Let $g$ be an $s$-sparse multivariate $GF(2)$ polynomial which does not contain the constant-1 term. Then $g(x) = 0$ for at least a $1/(s+1)$ fraction of all inputs.

Applying this corollary to the polynomial $p'_1$, we have that $p'_1$ is 0 on at least a $1/(s+1)$ fraction of its inputs. Therefore, the polynomial $T' + p_1$ is 1 on at least a $(\tau/s) \cdot 1/(s+1)$ fraction of all inputs in $\{0,1\}^n$; this in turn implies that $\mathrm{Vr}_p(i) \geq (\tau/s) \cdot 1/(s+1) = \tau/(s^2 + s)$. ∎

By a simple application of Lemma 5 we can show that setting low-variation variables to zero does not change the polynomial by much:

**Lemma 6.** *Let $p : \{0,1\}^n \to \{-1,1\}$ be an $s$-sparse polynomial. Let $g$ be a function obtained from $p$ by setting to 0 some subset of variables all of which have $\mathrm{Vr}_p(i) < \tau/(2s^2)$. Then $g$ and $p$ are $\tau$-close.*

*Proof.* Setting a variable to 0 removes all the terms that contain it from $p$. By Lemma 5, doing this only removes terms of length greater than $\log(s/\tau)$. Removing one such term changes the function on at most a $\tau/s$ fraction of the inputs. Since there are at most $s$ terms in total, the lemma follows by a union bound. ∎

### A.2 Partitioning variables into random subsets

The following lemma is at the heart of Theorem 3. The lemma states that when we randomly partition the variables (coordinates) into subsets, (*i*) each subset gets at most one "high-influence" variable (the term "high-influence" here means relative to an appropriate threshold value $t \ll \alpha$), and (*ii*) the remaining (low-influence) variables (w.r.t. $t$) have a "very small" contribution to the subset's total variation.

The first part of the lemma follows easily from a birthday–paradox type argument, since there are many more subsets than high-influence variables. As intuition for the second part, we note that in expectation, the total variation of each subset is very small. A more careful argument lets us argue that the total contribution of the low-influence variables in a given subset is unlikely to highly exceed its expectation.

**Lemma 7.** *Fix a value of $\alpha$ satisfying the first statement of Theorem 3. Let $t \stackrel{\text{def}}{=} \Delta\tau/(4C's)$, where $C'$ is a suitably large constant. Then with probability $99/100$ over the random partition the following statements hold true:*

- *For every $j \in [r]$, $I_j$ contains at most one variable $x_i$ with $\mathrm{Vr}_p(i) > t$.*
- *Let $I_j^{\leq t} \stackrel{\text{def}}{=} \{i \in I_j \mid \mathrm{Vr}_p(i) \leq t\}$. Then, for all $j \in [r]$, $\mathrm{Vr}_p(I_j^{\leq t}) \leq \Delta$.*

*Proof.* We show that each statement of the lemma fails independently with probability at most $1/200$ from which the lemma follows.

By Lemma 3 there are at most $b = s \log(2s/t)$ coordinates in $[n]$ with variation more than $t$. A standard argument yields that the probability there exists a subset $I_j$ with more than one such variable is at most $b^2/r$. It is easy to verify that this is less than $1/200$, as long as $C$ is large enough relative to $C'$. Therefore, with probability at least $199/200$, every subset contains at most one variable with variation greater than $t$. So the first statement fails with probability no more than $1/200$.

Now for the second statement. Consider a fixed subset $I_j$. We analyze the contribution of variables in $I_j^{\leq t}$ to the total variation $\mathrm{Vr}_p(I_j)$. We will show that with high probability the contribution of these variables is at most $\Delta$.

Let $S = \{i \in [n] \mid \mathrm{Vr}_p(i) \leq t\}$ and renumber the coordinates such that $S = [k']$. Each variable $x_i$, $i \in S$, is contained in $I_j$ independently with probability $1/r$. Let $X_1, \ldots, X_{k'}$ be the corresponding independent Bernoulli random variables. Recall that, by sub-additivity, the variation of $I_j^{\leq t}$ is upper bounded by $X = \sum_{i=1}^{k'} \mathrm{Vr}_p(i) \cdot X_i$. It thus suffices to upper bound the probability $\Pr[X > \Delta]$. Note that $\mathbb{E}[X] = \sum_{i=1}^{k'} \mathrm{Vr}_p(i) \cdot \mathbb{E}[X_i] = (1/r) \cdot \sum_{i=1}^{k'} \mathrm{Vr}_p(i) \leq (s/r)$, since $\sum_{i=1}^{k'} \mathrm{Vr}_p(i) \leq \sum_{i=1}^{n} \mathrm{Vr}_p(i) \leq s$. The last inequality follows from the following simple fact (the proof of which is left for the reader).

**Fact 4** *Let $p: \{0,1\}^n \to \{-1,1\}$ be an $s$-sparse polynomial. Then $\sum_{i=1}^{n} \mathrm{Vr}_p(i) \leq s$.*

To finish the proof, we need the following version of the Chernoff bound:

**Fact 5 ([MR95])** *For $k' \in \mathbb{N}^*$, let $\alpha_1, \ldots, \alpha_{k'} \in [0,1]$ and let $X_1, \ldots, X_{k'}$ be independent Bernoulli trials. Let $X' = \sum_{i=1}^{k'} \alpha_i X_i$ and $\mu \stackrel{\text{def}}{=} \mathbb{E}[X'] \geq 0$. Then for any $\gamma > 1$ we have $\Pr[X' > \gamma \cdot \mu] < (\frac{e^{\gamma-1}}{\gamma^\gamma})^\mu$.*

We apply the above bound for the $X_i$'s with $\alpha_i = \mathrm{Vr}_p(i)/t \in [0,1]$. (Recall that the coordinates in $S$ have variation at most $t$.) We have $\mu = \mathbb{E}[X'] = \mathbb{E}[X]/t \leq s/(rt) = C's/C\tau$, and we are interested in the event $\{X > \Delta\} \equiv \{X' > \Delta/t\}$. Note that $\Delta/t = 4C's/\tau$. Hence, $\gamma \geq 4C$ and the above bound implies that $\Pr[X > \Delta] < \left(e/(4C)\right)^{4C's/\tau} < (1/4C^4)^{C's/\tau}$.

Therefore, for a fixed subset $I_j$, we have $\Pr[\mathrm{Vr}_p(I_j^{\leq t}) > \Delta] < (1/4C^4)^{C's/\tau}$. By a union bound, we conclude that this happens in every subset with failure probability at most $r \cdot (1/4C^4)^{C's/\tau}$. This is less than $1/200$ as long as $C'$ is a large enough absolute constant (independent of $C$), which completes the proof. ∎

Next we show that by "zeroing out" the variables in low-variation subsets, we are likely to "kill" all terms in $p$ that contain a low-influence variable.

**Lemma 8.** *With probability at least $99/100$ over the random partition, every monomial of $p$ containing a variable with influence at most $\alpha$ has at least one of its variables in $\cup_{j \in L(\alpha)} I_j$.*

*Proof.* By Lemma 3 there are at most $b = s \log(8s^3/\tau)$ variables with influence more than $\alpha$. Thus, no matter the partition, at most $b$ subsets from $\{I_j\}_{j=1}^r$ contain such variables. Fix a low-influence variable (influence at most $\alpha$) from every monomial containing such a variable. For each fixed variable, the probability that it ends up in the same subset as a high-influence variable is at most $b/r$. Union bounding over each of the (at most $s$) monomials, the failure probability of the lemma is upper bounded by $sb/r < 1/100$. ∎

### A.3 Proof of Theorem 3

*Proof.* (Theorem 3) We prove each statement in turn. The first statement of the theorem is implied by Lemma 4. (Note that, as expected, the validity of this statement does not depend on the random partition.)

We claim that statements 2-5 essentially follow from Lemma 7. (In contrast, the validity of these statements crucially depends on the random partition.)

Let us first prove the third statement. We want to show that (w.h.p. over the choice of $\alpha$ and $\{I_j\}_{j=1}^r$) for every $j \in H(\alpha)$, (*i*) there exists a *unique* $i_j \in I_j$ such that $\mathrm{Vr}_p(i_j) \geq \alpha$ and (*ii*) that $\mathrm{Vr}_p(I_j \setminus \{i_j\}) \leq \Delta$. Fix some $j \in H(\alpha)$. By Lemma 7, for a given value of $\alpha$ satisfying the first statement of the theorem, we have: (*i'*) $I_j$ contains at most one variable $x_{i_j}$ with $\mathrm{Vr}_p(i_j) > t$ and (*ii'*) $\mathrm{Vr}_p(I_j \setminus \{i_j\}) \leq \Delta$. Since $t < \tau/4s^2 < \alpha$ (with probability 1), (*i'*) clearly implies that, if $I_j$ has a high-variation element (w.r.t. $\alpha$), then it is unique. In fact, we claim that $\mathrm{Vr}_p(i_j) \geq \alpha$. For otherwise, by sub-additivity of variation, we would have $\mathrm{Vr}_p(I_j) \leq \mathrm{Vr}_p(I_j \setminus \{i_j\}) + \mathrm{Vr}_p(i_j) \leq \Delta + \alpha - 4\Delta = \alpha - 3\Delta < \alpha$, which contradicts the assumption that $j \in H(\alpha)$. Note that we have used the fact that $\alpha$ satisfies the first statement of the theorem, that is $\mathrm{Vr}_p(i_j) < \alpha \Rightarrow \mathrm{Vr}_p(i_j) < \alpha - 4\Delta$. Hence, for a "good" value of $\alpha$ (one satisfying the first statement of the theorem), the third statement is satisfied with probability at least $99/100$ over the random partition. By Lemma 4, a "good" value of $\alpha$ is chosen with probability $96/100$. By independence, the conclusions of Lemma 4 and Lemma 7 hold simultaneously with probability more than $9/10$.

We now establish the second statement. We assume as before that $\alpha$ is a "good" value. Consider a fixed subset $I_j$, $j \in [r]$. If $j \in H(\alpha)$ (i.e. $I_j$ is a high-variation subset) then, with probability at least $99/100$ (over the random partition), there exists $i_j \in I_j$ such that $\mathrm{Vr}_p(i_j) \geq \alpha + 4\Delta$. The monotonicity of variation yields $\mathrm{Vr}_p(I_j) \geq \mathrm{Vr}_p(i_j) \geq \alpha + 4\Delta$. If $j \in L(\alpha)$ then $I_j$ contains no high-variation variable, i.e. its maximum variation element has variation at most $\alpha - 4\Delta$ and by the second part of Lemma 7 the remaining variables contribute at most $\Delta$ to its total variation. Hence, by sub-additivity we have that $\mathrm{Vr}_p(I_j) \leq \alpha - 3\Delta$. Since a "good" value of $\alpha$ is chosen with probability $96/100$, the desired statement follows.

The fourth statement follows from the aforementioned and the fact that there exist at most $s \log(8s^3/\tau)$ variables with variation at least $\alpha$ (as follows from Lemma 3, given that $\alpha > \tau/(4s^2)$).

Now for the fifth statement. Lemma 8 and monotonicity imply that the only variables that remain relevant in $p'$ are (some of) those with high influence (at least $\alpha$) in $p$, and, as argued above, each high-variation subset $I_j$ contains at most one such variable.

By a union bound, the conclusion of Lemma 8 holds simultaneously with the conclusions of Lemma 4 and Lemma 7 with probability at least $9/10$.

The sixth statement (that $p$ and $p'$ are $\tau$-close) is a consequence of Lemma 6 (since $p'$ is obtained from $p$ by setting to 0 variables with variation less than $\alpha < \tau/(2s^2)$). This concludes the proof of Theorem 3. ∎

## B  Completeness of the test

In this section we show that **Test-Sparse-Poly** is complete:

**Theorem 6.** *Suppose $f$ is an $s$-sparse $GF(2)$ polynomial. Then **Test-Sparse-Poly** accepts $f$ with probability at least $2/3$.*

*Proof.* Fix $f$ to be an $s$-sparse $GF(2)$ polynomial over $\{0,1\}^n$. By the choice of the $\Delta$ and $r$ parameters in Step 1 of **Test-Sparse-Poly** we may apply Theorem 3, so with failure probability at most $1/10$ over the choice of $\alpha$ and $I_1, \ldots, I_r$ in Steps 2 and 3, statements 1–6 of Theorem 3 all hold. We shall write $f'$ to denote $f|_{0 \leftarrow \cup_{j \in L(\alpha)} I_j}$. Note that at each successive stage of the proof we shall assume that the "failure probability" events do not occur, i.e. henceforth we shall assume that statements 1–6 all hold for $f$; we take a union bound over all failure probabilities at the end of the proof.

Now consider the $M$ executions of the independence test for a given fixed $I_j$ in Step 4. Lemma 2 gives that each run rejects with probability $\frac{1}{2}\mathrm{Vr}_f(I_j)$. A standard Hoeffding bound implies that for the algorithm's choice of $M = \frac{2}{\Delta^2}\ln(200r)$, the value $\widetilde{\mathrm{Vr}}_f(I_j)$ obtained in Step 4 is within $\pm\Delta$ of the true value $\mathrm{Vr}_f(I_j)$ with failure probability at most $\frac{1}{100r}$. A union bound over all $j \in [r]$ gives that with failure probability at most $1/100$, we have that each $\widetilde{\mathrm{Vr}}_f(I_j)$ is within an additive $\pm\Delta$ of the true value $\mathrm{Vr}_f(I_j)$. This means that (by statement 2 of Theorem 3) every $I_j$ has $\widetilde{\mathrm{Vr}}_f(I_j) \notin [\alpha - 2\Delta, \alpha + 3\Delta]$, and hence in Step 5 of the test, the sets $\widetilde{L}(\alpha)$ and $\widetilde{H}(\alpha)$ are identical to $L(\alpha)$ and $H(\alpha)$ respectively, which in turn means that the function $\widetilde{f'}$ defined in Step 5 is identical to $f'$ defined above.

We now turn to Step 6 of the test. By statement 6 of Theorem 3 we have that $f$ and $f'$ disagree on at most a $\tau$ fraction of inputs. A union bound over the $m$ random examples drawn in Step 6 implies that with failure probability at most $\tau m < 1/100$ the test proceeds to Step 7.

By statement 3 of Theorem 3 we have that each $I_j$, $j \in \widetilde{H}(\alpha) \equiv H(\alpha)$, contains precisely one high-variation element $i_j$ (i.e. which satisfies $\mathrm{Vr}_f(i_j) \geq \alpha$), and these are all of the high-variation elements. Consider the set of these $|\widetilde{H}(\alpha)|$ high-variation variables; statement 5 of Theorem 3 implies that these are the only variables which $f'$ can depend on (it is possible that it does not depend on some of these variables). Let us write $f''$ to denote the function $f'' : \{0,1\}^{|\widetilde{H}(\alpha)|} \to \{-1,1\}$ corresponding to $f'$ but whose input variables are these $|\widetilde{H}(\alpha)|$ high-variation variables in $f$, one per $I_j$ for each $j \in \widetilde{H}(\alpha)$. We thus have that $f''$ is isomorphic to $f'$ (obtained from $f'$ by discarding irrelevant variables).

The main idea behind the completeness proof is that in Step 7 of **Test-Sparse-Poly**, the learning algorithm **LearnPoly**′ is being run with target function $f''$. Since $f''$ is isomorphic to $f'$, which is an $s$-sparse polynomial (since it is a restriction of an $s$-sparse polynomial $f$), with high probability **LearnPoly'** will run successfully and the test will accept. To show that this is what actually happens, we must show that with high probability each call to **SimMQ** which **LearnPoly'** makes correctly simulates the corresponding membership query to $f''$. This is established by the following lemmas:

**Lemma 9.** *Let $f, I, \alpha, \Delta$ be such that $I$ is $(\alpha, \Delta)$-well-structured with $\Delta \leq \alpha\delta/(2\ln(2/\delta))$. Then with probability at least $1 - \delta$, the output of* **SHIV**$(f, I, \alpha, \Delta, b, \delta)$ *is an assignment $w \in \{0,1\}^I$ which has $w_i = b$.*

*Proof.* We assume that $I^b$ contains the high-variation variable $i$ (the other case being very similar). Recall that by Lemma 2, each run of the independence test on $I^b$ rejects with probability $\frac{1}{2}\mathrm{Vr}_f(I^b)$; by Lemma 1 (monotonicity) this is at least $\frac{1}{2}\mathrm{Vr}_f(i) \geq \alpha/2$. So the probability that $I^b$ is not marked even once after $c$ iterations of the independence test is at most $(1 - \alpha/2)^c \leq \delta/2$, by our choice of $c$. Similarly, the probability that $I^{\overline{b}}$ is ever marked during $c$ iterations of the independence test is at most $c(\Delta/2) \leq \delta/2$, by the condition of the lemma. Thus, the probability of failing at step 3 of **SHIV** is at most $\delta$, and since $i \in I^b$, the assignment $w$ sets variable $i$ correctly in step 4. ∎

**Lemma 10.** *With total failure probability at most $1/100$, each of the $Q(s, |\widetilde{H}(\alpha)|, \epsilon/4, 1/100)$ calls to* **SimMQ**$(f, \widetilde{H}(\alpha), \{I_j\}_{j \in \widetilde{H}(\alpha)}, \alpha, \Delta, z, 1/(100Q(s, |\widetilde{H}(\alpha)|, \epsilon/4, 1/100)))$ *that* **LearnPoly'** *makes in Step 7 of* **Test-Sparse-Poly** *returns the correct value of $f''(z)$.*

*Proof.* Consider a single call to the procedure **SimMQ**$(f, \widetilde{H}(\alpha), \{I_j\}_{j \in \widetilde{H}(\alpha)}, \alpha, \Delta, z, 1/(100Q(s, |\widetilde{H}(\alpha)|, \epsilon/4, 1/100)))$ made by **LearnPoly'**. We show that with failure probability at most $\delta' \stackrel{\text{def}}{=} 1/(100Q(s, |\widetilde{H}(\alpha)|, \epsilon/4, 1/100))$ this call returns the value $f''(z)$, and the lemma then follows by a union bound over the $Q(s, |\widetilde{H}(\alpha)|, \epsilon/4, 1/100)$ many calls to **SimMQ**.

This call to **SimMQ** makes $|\widetilde{H}(\alpha)|$ calls to **SHIV**$(f, I_j, \alpha, \Delta, z_j, \delta'/|\widetilde{H}(\alpha)|)$, one for each $j \in \widetilde{H}(\alpha)$. Consider any fixed $j \in \widetilde{H}(\alpha)$. Statement 3 of Theorem 3 gives that $I_j$ ($j \in \widetilde{H}(\alpha)$) is $(\alpha, \Delta)$-well-structured. Since $\alpha > \frac{\tau}{4s^2}$, it is easy to check the condition of Lemma 9 holds where the role of "$\delta$" in that inequality is played by $\delta'/|\widetilde{H}(\alpha)|$, so we may apply Lemma 9 and conclude that with failure probability at most $\delta'/|\widetilde{H}(\alpha)|$ (recall that by statement 4 of Theorem 3 we have $|\widetilde{H}(\alpha)| \leq s\log(8s^3/\tau)$), **SHIV** returns an assignment to the variables in $I_j$ which sets the high-variation variable to $z_j$ as required. By a union bound, the overall failure probability that any $I_j$ ($j \in \widetilde{H}(\alpha)$) has its high-variation variable not set according to $z$ is at most $\delta'$. Now statement 5 and the discussion preceding this lemma (the isomorphism between $f'$ and $f''$) give that **SimMQ** sets all of the variables that are relevant in $f'$ correctly according to $z$ in the assignment $x$ it constructs in Step 2. Since this assignment $x$ sets all variables in $\cup_{j \in \widetilde{L}} I_j$ to 0, the bit $b = f(x)$ that is returned is the correct value of $f''(z)$, with failure probability at most $\delta'$ as required. ∎

With Lemma 10 in hand, we have that with failure probability at most $1/100$, the execution of **LearnPoly'**$(s, |\widetilde{H}(\alpha)|, \epsilon/4, 1/100)$ in Step 7 of **Test-Sparse-Poly** correctly simulates all membership queries. As a consequence, Corollary 1 thus gives

that **LearnPoly'**$(s, |\widetilde{H}(\alpha)|, \epsilon/4, 1/100))$ returns "not $s$-sparse" with probability at most $1/100$. Summing all the failure probabilities over the entire execution of the algorithm, the overall probability that **Test-Sparse-Poly** does not output "yes" is at most

$$\overbrace{1/10}^{\text{Theorem 3}} + \overbrace{1/100}^{\text{Step 4}} + \overbrace{1/100}^{\text{Step 6}} + \overbrace{1/100}^{\text{Lemma 10}} + \overbrace{1/100}^{\text{Corollary 1}} < 1/5,$$

and the completeness theorem is proved. (Theorem 6) ∎

## C  Soundness of the Test

In this section we prove the soundness of **Test-Sparse-Poly**:

**Theorem 7.** *If $f$ is $\epsilon$-far from any $s$-sparse polynomial, then **Test-Sparse-Poly** accepts with probability at most $1/3$.*

*Proof.* To prove the soundness of the test, we start by assuming that the function $f$ has progressed to step 5, so there are subsets $I_1, \ldots, I_r$ and $\widetilde{H}(\alpha)$ satisfying $\widetilde{\mathrm{Vr}}_f(I_j) > \alpha + 2\Delta$ for all $j \in \widetilde{H}(\alpha)$. As in the proof of completeness, we have that the actual variations of all subsets should be close to the estimates, i.e. that $\mathrm{Vr}_f(I_j) > \alpha + \Delta$ for all $j \in \widetilde{H}(\alpha)$ except with with probability at most $1/100$. We may then complete the proof in two parts by establishing the following:

- If $f$ and $\widetilde{f'}$ are $\epsilon_a$-far, step 6 will accept with probability at most $\delta_a$.
- If $\widetilde{f'}$ is $\epsilon_b$-far from every $s$-sparse polynomial, step 7 will accept with probability at most $\delta_b$.

Establishing these statements with $\epsilon_a = \epsilon_b = \epsilon/2$, $\delta_a = 1/12$ and $\delta_b = 1/6$ will allow us to complete the proof (and we may assume throughout the rest of the proof that $\mathrm{Vr}_f(I_j) > \alpha$ for each $j \in \widetilde{H}(\alpha)$).

The first statement follows immediately by our choice of $m = \frac{1}{\epsilon_a} \ln \frac{1}{\delta_a}$ with $\epsilon_a = \epsilon/2$ and $\delta_a = 1/12$ in Step 6. Our main task is to establish the second statement, which we do using Lemmas 11 and 12 stated below. Intuitively, we would like to show that if **LearnPoly'** outputs a hypothesis $h$ (which must be an $s$-sparse polynomial since **LearnPoly'** is proper) with probability greater than $1/6$, then $\widetilde{f'}$ is close to a junta isomorphic to $h$. To do this, we establish that if **LearnPoly'** succeeds with high probability, then the last hypothesis on which an equivalence query is performed in **LearnPoly'** is a function which is close to $\widetilde{f'}$. Our proof uses two lemmas: Lemma 12 tells us that this holds if the high variation subsets satisfy a certain structure, and Lemma 11 tells us that if **LearnPoly'** succeeds with high probability then the subsets indeed satisfy this structure. We now state these lemmas formally and complete the proof of the theorem, deferring the proofs of the lemmas until later.

Recall that the algorithm **LearnPoly'** will make repeated calls to **SimMQ** which in turn makes repeated calls to **SHIV**. Lemma 11 states that if, with probability greater than $\delta_2$, all of these calls to **SHIV** return without failure, then the subsets associated with $\widetilde{H}(\alpha)$ have a special structure.

**Lemma 11.** *Let $J \subset [n]$ be a subset of variables obtained by including the highest-variation element in $I_j$ for each $j \in \widetilde{H}(\alpha)$ (breaking ties arbitrarily). Suppose that $k > 300|\widetilde{H}(\alpha)|/\epsilon_2$ queries are made to **SimMQ**. Suppose moreover that $\Pr[$ every call to **SHIV** that is made during these $k$ queries returns without outputting 'fail'$]$ is greater than $\delta_2$ for $\delta_2 = 1/\Omega(k)$. Then the following both hold:*

- *Every subset $I_j$ for $j \in \widetilde{H}(\alpha)$ satisfies $\mathrm{Vr}_f(I_j \setminus J) \leq 2\epsilon_2/|\widetilde{H}(\alpha)|$; and*
- *The function $\widetilde{f}'$ is $\epsilon_2$-close to the junta $g : \{0,1\}^{|\widetilde{H}(\alpha)|} \to \{-1,1\}$ defined as as:*

$$g(x) \stackrel{\text{def}}{=} \mathrm{sign}(\mathbb{E}_z[\widetilde{f}'((x \cap J) \sqcup z)]).$$

Given that the subsets associated with $\widetilde{H}(\alpha)$ have this special structure, Lemma 12 tells us that the hypothesis output by **LearnPoly**$'$ should be close to the junta $g$.

**Lemma 12.** *Define $Q_E$ as the maximum number of calls to **SimMQ** that that will be made by **LearnPoly**$'$ in all of its equivalence queries. Suppose that for every $j \in \widetilde{H}(\alpha)$, it holds that $\mathrm{Vr}_f(I_j \setminus J) < 2\epsilon_2/|\widetilde{H}(\alpha)|$ with $\epsilon_2 < \frac{\alpha}{800 Q_E}$. Then the probability that **LearnPoly**$'$ outputs a hypothesis $h$ which is $\epsilon/4$-far from the junta $g$ is at most $\delta_3 = 1/100$.*

We now show that Lemmas 11 and 12 suffice to prove the desired result. Suppose that **LearnPoly**$'$ accepts with probability at least $\delta_b = 1/6$. Assume **LearnPoly**$'$ makes at least $k$ queries to **SimMQ** (we address this in the next paragraph); then it follows from Lemma 11 that the bins associated with $\widetilde{H}(\alpha)$ satisfy the conditions of Lemma 12 and that $\widetilde{f}'$ is $\epsilon_2$-close to the junta $g$. Now applying Lemma 12, we have that with failure probability at most $1/100$, **LearnPoly**$'$ outputs a hypothesis which is $\epsilon/4$-close to $g$. But then $\widetilde{f}'$ must be $(\epsilon_2 + \epsilon/4)$-close to this hypothesis, which is an $s$-sparse polynomial.

We need to establish that **LearnPoly**$'$ indeed makes $k > 300|\widetilde{H}(\alpha)|/\epsilon_2$ **SimMQ** queries for an $\epsilon_2$ that satisfies the condition on $\epsilon_2$ in Lemma 12. (Note that if **LearnPoly**$'$ does not actually make this many queries, we can simply have it make artificial calls to **SHIV** to achieve this. An easy extension of our completeness proof handles this slight extension of the algorithm; we omit the details.) Since we need $\epsilon_2 < \alpha/800 Q_E$ and Theorem 2 gives us that $Q_E = (|\widetilde{H}(\alpha)|s+2) \cdot \frac{4}{\epsilon} \ln 300(|\widetilde{H}(\alpha)|s+2)$ (each equivalence query is simulated using $\frac{4}{\epsilon} \ln 300(|\widetilde{H}(\alpha)|s+2)$ random examples), an easy computation shows that it suffices to take $k = \mathrm{poly}(s, 1/\epsilon)$, and the proof of Theorem 7 is complete. ∎

Before proving Lemma 12 and Lemma 11, we prove the following about the behavior of **SHIV** when it is called with parameters $\alpha, \Delta$ that do not quite match the real values $\alpha', \Delta'$ for which $I$ is $(\alpha', \Delta')$-well-structured:

**Lemma 13.** *If $I$ is $(\alpha', \Delta')$-well-structured, then the probability that $\mathbf{SHIV}(f, I, \alpha, \Delta, b, \delta)$ passes (i.e. does not output "fail") and sets the high variation variable incorrectly is at most $(\delta/2)^{\alpha'/\alpha} \cdot (1/\alpha) \cdot \Delta' \cdot \ln(2/\delta)$.*

*Proof.* The only way for **SHIV** to pass with an incorrect setting of the high-variation variable $i$ is if it fails to mark the subset containing $i$ for $c$ iterations of the independence test, and marks the other subset at least once. Since $Vr(i) > \alpha'$ and $Vr(I \setminus i) < \Delta'$, the probability of this occurring is at most $(1 - \alpha'/2)^c \cdot \Delta' \cdot c/2$. Since **SHIV** is called with failure parameter $\delta$, $c$ is set to $\frac{2}{\alpha} \ln \frac{2}{\delta}$. ∎

We now give a proof of Lemma 12, followed by a proof of Lemma 11.

*Proof.* (Lemma 12) By assumption each $\mathrm{Vr}_f(I_j \setminus J) \leq 2\epsilon_2/|\widetilde{H}(\alpha)|$ and $\mathrm{Vr}_f(I_j) > \alpha$, so subadditivity of variation gives us that for each $j \in \widetilde{H}(\alpha)$, there exists an $i \in I_j$ such that $\mathrm{Vr}_f(i) > \alpha - 2\epsilon_2/|\widetilde{H}(\alpha)|$. Thus for every each call to **SHIV** made by **SimMQ**, the conditions of Lemma 13 are satisfied with $\mathrm{Vr}_f(i) > \alpha - 2\epsilon_2/|\widetilde{H}(\alpha)|$ and $\mathrm{Vr}_f(I_j \setminus J) < 2\epsilon_2/|\widetilde{H}(\alpha)|$. We show that as long as $\epsilon_2 < \frac{\alpha}{800 Q_E}$, the probability that any particular query $z$ to **SimMQ** has a variable set incorrectly is at most $\delta_3/3Q_E$.

Suppose **SHIV** has been called with failure probability $\delta_4$, then the probability given by Lemma 13 is at most:

$$(\delta_4/2)^{1-2\epsilon_2/(\alpha \cdot |\widetilde{H}(\alpha)|)} \cdot \frac{2}{\alpha} \ln\left(\frac{2}{\delta_4}\right) \cdot 2\epsilon_2/|\widetilde{H}(\alpha)|, \tag{1}$$

We shall show that this is at most $\delta_3/3|\widetilde{H}(\alpha)|Q_E = 1/300 Q_E|\widetilde{H}(\alpha)|$. Taking $\epsilon_2 \leq \alpha/800 Q_E$ simplifies (1) to:

$$\frac{1}{300 Q_E |\widetilde{H}(\alpha)|} \cdot (\delta_4/2)^{1-2\epsilon_2/(\alpha \cdot |\widetilde{H}(\alpha)|)} \cdot \frac{3}{4} \ln \frac{2}{\delta_4},$$

which is at most $1/300|\widetilde{H}(\alpha)|Q_E$ as long as

$$(2/\delta_4)^{1-2\epsilon_2/(\alpha \cdot |\widetilde{H}(\alpha)|)} > \frac{3}{4} \ln \frac{2}{\delta_4},$$

which certainly holds for our choice of $\epsilon_2$ and the setting of $\delta_4 = 1/100k|\widetilde{H}(\alpha)|$. Each call to **SimMQ** uses $|\widetilde{H}(\alpha)|$ calls to **SHIV**, so a union bound gives that each random query to **SimMQ** returns an incorrect assignment with probability at most $1/300Q_E$.

Now, since $\widetilde{f}'$ and $g$ are $\epsilon_2$-close and $\epsilon_2$ satisfies $\epsilon_2 Q_E \leq \delta_3/3$, in the uniform random samples used to simulate the final (accepting) equivalence query, **LearnPoly'** will receive examples labeled correctly according to $g$ with probability at least $1 - 2\delta_3/3$. Finally, note that **LearnPoly'** makes at most $|\widetilde{H}(\alpha)|s+2$ equivalence queries and hence each query is simulated using $\frac{4}{\epsilon} \ln \frac{3(|\widetilde{H}(\alpha)|s+2)}{\delta_3}$ random examples (for a failure probability of $\frac{\delta_3}{|\widetilde{H}(\alpha)|s+2}$ for each equivalence query). Then **LearnPoly'** will reject with probability at least $1 - \delta_3/3$ unless $g$ and $h$ are $\epsilon/4$-close. This concludes the proof of Lemma 12. ∎

*Proof.* (Lemma 11) We prove that if $\mathrm{Vr}_f(I_j \setminus J) > 2\epsilon_2/|\widetilde{H}(\alpha)|$ for some $j \in \widetilde{H}(\alpha)$, then the probability that all calls to **SHIV** return successfully is at most $\delta_2$. The closeness of $\widetilde{f}'$ and $g$ follows easily by the subadditivity of variation and Proposition 3.2 of [FKR+04].

First, we prove a much weaker statement whose analysis and conclusion will be used to prove the proposition. We show in Proposition 1 that if the test accepts with high probability, then the variation from each variable in any subset is small. We use the bound on each variable's variation to obtain the concentration result in Proposition 2, and then complete the proof of Lemma 11.

**Proposition 1.** *Suppose that $k$ calls to **SHIV** are made with a particular subset $I$, and let $i$ be the variable with the highest variation in $I$. If $\mathrm{Vr}_f(j) > \epsilon_2/100|\widetilde{H}(\alpha)|$ for some $j \in I \setminus i$, then the probability that **SHIV** returns without outputting 'fail' for all $k$ calls is at most $\delta^* = e^{-k/18} + e^{-c}$.*

*Proof.* Suppose that there exist $j, j' \in I$ with $\mathrm{Vr}_f(j) \geq \mathrm{Vr}_f(j') \geq \epsilon_2/100|\widetilde{H}(\alpha)|$. A standard Chernoff bound gives that except with probability at most $e^{-k/18}$, for at least $(1/3)k$ of the calls to **SHIV**, variables $j$ and $j'$ are in different partitions. In these cases, the probability **SHIV** does not output 'fail' is at most $2(1 - \epsilon_2/100|\widetilde{H}(\alpha)|)^c$, since for each of the $c$ runs of the independence test, one of the partitions must not be marked. The probability no call outputs 'fail' is at most $e^{-k/18} + 2(1 - \epsilon_2/100|\widetilde{H}(\alpha)|)^{ck/3}$. Our choice of $k > 300|\widetilde{H}(\alpha)|/\epsilon_2$ ensures that $(1/e)^{ck\epsilon_2/300|\widetilde{H}(\alpha)|} \leq (1/e)^c$. ■

Since in our setting $|I_j|$ may depend on $n$, using the monotonicity of variation with the previous claim does not give a useful bound on $\mathrm{Vr}_f(I \setminus i)$. But we see from the proof that if the variation of each partition is not much less than $\mathrm{Vr}_f(I \setminus i)$ and $\mathrm{Vr}_f(I \setminus i) > 2\epsilon_2/|\widetilde{H}(\alpha)|$, then with enough calls to **SHIV** one of these calls should output "fail." Hence the lemma will be easily proven once we establish the following proposition:

**Proposition 2.** *Suppose that $k$ calls to **SHIV** are made with a particular subset $I$ having $\mathrm{Vr}_f(I \setminus i) > 2\epsilon_2/|\widetilde{H}(\alpha)|$ and $\mathrm{Vr}_f(j) \leq \epsilon_2/100|\widetilde{H}(\alpha)|$ for every $j \in I \setminus i$. Then with probability greater than $1 - \delta^{**} = 1 - e^{-k/18}$, at least $1/3$ of the $k$ calls to **SHIV** yield both $\mathrm{Vr}_f(I^1) > \eta \mathrm{Vr}_f(I \setminus i)/2$ and $\mathrm{Vr}_f(I^0) > \eta \mathrm{Vr}_f(I \setminus i)/2$, where $\eta = 1/e - 1/50$.*

*Proof.* We would like to show that a random partition of $I$ into two parts will result in parts each of which has variation not much less than the variation of $I \setminus i$. Choosing a partition is equivalent to choosing a random subset $I'$ of $I \setminus i$ and including $i$ in $I'$ or $I \setminus I'$ with equal probability. Thus it suffices to show that for random $I' \subseteq I \setminus i$, it is unlikely that $\mathrm{Vr}_f(I')$ is much smaller than $\mathrm{Vr}_f(I \setminus i)$.

This does not hold for general $I$, but by bounding the variation of any particular variable in $I$, which we have done in Proposition 1, and computing the *unique-variation* (a technical tool introduced in [FKR$^+$04]) of $I'$, we may obtain a deviation bound on $\mathrm{Vr}_f(I')$. The following statement follows from Lemma 3.4 of [FKR$^+$04]:

**Proposition 3 ([FKR$^+$04]).** *Define the unique-variation of variable $j$ (with respect to $i$) as*

$$\mathrm{Ur}_f(j) = \mathrm{Vr}_f([j] \setminus i) - \mathrm{Vr}_f([j-1] \setminus i).$$

*Then for any $I' \subseteq I \setminus i$,*

$$\mathrm{Vr}_f(I') \geq \sum_{j \in I'} \mathrm{Ur}_f(j) = \sum_{j \in I'} \mathrm{Vr}_f([j] \setminus i) - \mathrm{Vr}_f([j-1] \setminus i).$$

Now $\mathrm{Vr}_f(I')$ is lower bounded by a sum of independent, non-negative random variables whose expectation is given by

$$\mathbb{E}[\sum_{j \in I'} \mathrm{Ur}_f(j)] = \sum_{j=1}^{n} (1/2)\mathrm{Ur}_f(j) = \mathrm{Vr}_f(I \setminus i)/2 \stackrel{\mathrm{def}}{=} \mu.$$

To obtain a concentration property, we require a bound on each $\mathrm{Ur}_f(j) \leq \mathrm{Vr}_f(j)$, which is precisely what we showed in the previous proposition. Note that $\mathrm{Ur}_f(i) = 0$, and recall that we have assumed that $\mu > \epsilon_2/|\widetilde{H}(\alpha)|$ and every $j \in I \setminus i$ satisfies $\mathrm{Vr}_f(j) < \mu/100$.

Now we may use the bound from [FKR+04] in Proposition 3.5 with $\eta = 1/e - 2/100$ to obtain:

$$\Pr[\sum_{j \in I'} \mathrm{Ur}_f(j) < \eta\mu] < \exp(\frac{100}{e}(\eta e - 1)))] \leq 1/e^2.$$

Thus the probability that one of $I^0$ and $I^1$ has variation less than $\eta\mu$ is at most $1/2$. We expect that half of the $k$ calls to **SHIV** will result in $I^0$ and $I^1$ having variation at least $\eta\mu$, so a Chernoff bound completes the proof of the claim with $\delta^{**} \leq e^{-k/18}$. This concludes the proof of Proposition 2. ∎

Finally, we proceed to prove the lemma. Suppose that there exists some $I$ such that $\mathrm{Vr}_f(I \setminus i) > 2\epsilon_2/|\widetilde{H}(\alpha)|$. Now the probability that a particular call to **SHIV** with subset $I$ succeeds is:

$$\Pr[\mathrm{marked}(I^0); \neg \mathrm{marked}(I^1)] + \Pr[\mathrm{marked}(I^1); \neg \mathrm{marked}(I^0)].$$

By Propositions 1 and 2, if with probability at least $\delta^* + \delta^{**}$ none of the $k$ calls to **SHIV** return fail, then for $k/3$ runs of **SHIV** both $\mathrm{Vr}_f(I^1)$ and $\mathrm{Vr}_f(I^0)$ are at least $\eta\epsilon_2/|\widetilde{H}(\alpha)| > \epsilon_2/4|\widetilde{H}(\alpha)|$ and thus both probabilities are at most $(1 - \epsilon_2/4|\widetilde{H}(\alpha)|)^c$.

As in the analysis of the first proposition, we may conclude that every subset $I$ which is called with **SHIV** at least $k$ times either satisfies $\mathrm{Vr}_f(I \setminus i) < 2\epsilon_2/|\widetilde{H}(\alpha)|$ or will cause the test to reject with probability at least $1 - \delta^{**} - 2\delta^*$. Recall that $\delta^* = e^{-c} + e^{-k/18}$; since **SHIV** is set to run with failure probability at most $1/|\widetilde{H}(\alpha)|k$, we have that $\delta_2$ is $1/\Omega(k)$. This concludes the proof of Lemma 11. ∎